\documentclass[]{aastex7}
\usepackage{amsmath}
\usepackage{soul}
\usepackage{hyperref}

\newcommand{\kms}{km s$^{-1}$}
\newcommand{\me}{\mathrm{e}}
\newcommand{\be}{\begin{equation}}
\newcommand{\ee}{\end{equation}}

\submitjournal{ApJ}

\begin{document}

\title{First Detection of Circular Polarization in 4.7 GHz Excited OH Masers}
\shorttitle{Circular Polarization in 4.7 GHz Masers}

\author[0000-0003-3593-9707]{Derck P. Smits}
\affiliation{Dept of Mathematical Sciences, University of South Africa, Private 
Bag X6, Florida, 1709, South Africa}
\email{derck.smits@gmail.com}

\author[0000-0002-4846-1741]{Paul Fallon}
\affiliation{Centre for Space Research, North-West University, Private Bag X1290, 
Potchefstroom 2520, South Africa}
\affiliation{Dept of Mathematical Sciences, University of South Africa, Private 
Bag X6, Florida, 1709, South Africa}
\email{paulfallon@telkomsa.net}
\shortauthors{Smits \& Fallon}

\correspondingauthor{Derck P. Smits}

\begin{abstract}

We report the first detection of circular polarization in 4.7 GHz excited 
OH masers in star-forming regions made using full Stokes measurements with 
the Green Bank 100m telescope. The Zeeman shift between the two circular 
components provides a measure of the magnetic field pervading these maser 
spots. Three different methods are used to determine the shift in velocity 
between the RCP and LCP components. We find fields with $B \sim 100$ mG 
using archival molecular parameters that have limited precision and 
uncertain values. Reservations of using 1.7 and 6.0 GHz OH masers to 
estimate magnetic fields in star-forming regions are discussed.
\end{abstract}

\keywords{Hydroxyl masers (772) --- Interstellar magnetic fields (845) --- 
Spectropolarimetry (1973) --- Young stellar objects (1834)}

\section{Introduction} \label{sec:intro}

Astronomical masers were discovered about 60 years ago when \citet{WMC65} 
recognized that the narrow lines of `mysterium' suggested by \citet{WWD65} 
corresponded to the 1.665 GHz groundstate (gs) OH transition. Their narrow
widths identified them as maser rather than thermal emission. Soon after
this, lines from excited (ex) OH at 4.7 GHz and 6.0 GHz were also
discovered by \citet{ZPP68} and \citet{YZP69}, respectively. In all
instances, these masers occurred in molecular clouds associated with
star-forming regions. Many other species of masers have been detected
subsequently in a variety of astronomical environments. 

An interesting feature of all the maser lines from 1.7 GHz gsOH and the 
6.0 GHz exOH lines in these discovery observations was that they were 
circularly polarized. These levels are both from the $^2\Pi_{3/2}$ chain 
of the OH energy levels, for which it was assumed that this polarization 
was due to the presence of magnetic fields in the molecular clouds 
causing Zeeman splitting of the levels of OH. For both species of masers, 
isolated maser spots of either RCP or LCP emission have been found. 
These individual spots provide no measure of the magnetic field, only 
spatially-related pairs of spots can be used. In contrast to this, the 
4.7 GHz exOH masers were not polarized. However, these masers occur 
between states in the lowest level of the $^2\Pi_{1/2}$ chain which 
have a Land\'{e} g-factor about three orders of magnitude smaller than 
for the $^2\Pi_{3/2}$ levels, so that Zeeman splitting was neither 
expected nor observed.

Theoretical calculations by \citet{R61} that were compared with measured 
Zeeman splitting in OH laboratory experiments showed excellent agreement,
giving the Land\'{e} $g_{J}$ values for these levels as $g_J = 0.935$ 
and $0.485$ for the 1.7 and 6.0 GHz lines respectively. These parameters 
produce Zeeman splittings of 590 and 236 km s$^{-1}$ G$^{-1}$ in the 
1.665 and 1.720 GHz gsOH lines, and 79.0 and 56.4 km s$^{-1}$ G$^{-1}$ 
in the 6.031 and 6.035 GHz exOH lines (see Table 1 in \citet{D74}). 
Note that \citet{D74} quotes values for the separation between the 
RCP and LCP components which is twice the value of the velocity shift
due to the Zeeman effect. Based on the calculations of \citet{R61}, a 
value for the 4.7 GHz lines is given only as $\sim 10^{-2}$ km s$^{-1}$ 
G$^{-1}$. Clearly, the Zeeman splitting of these exOH lines is much 
smaller than those of lines in the $^2\Pi_{3/2}$ chain. 

Zeeman pairs in gsOH 1.720 GHz masers were definitively identified for 
the first time by \citet{LWB75}, using three station VLBI measurements to 
show the spatial correspondence of RCP and LCP pairs. Since then 1.7 and 
6.0 GHz gs- and exOH masers have regularly been used to determine the 
strength of magnetic fields in star-forming regions. Results from these 
observations indicate magnetic fields in the range of $1 - 10$ mG. 

The spectra of 18 cm gsOH masers are often complex and have overlapping 
components in single-dish measurements. Zeeman components can usually 
only be identified using array measurements showing maps of the maser 
spots and identifying RCP and LCP components that lie spatially near 
each other. Often there is less confusion in 6 GHz exOH masers but even 
so maps of the regions provide a more secure identification of Zeeman 
pairs than single-dish spectra, although there are problems with this 
approach as well. Orthogonally polarized pairs of spots with similar 
velocities are assumed to be at the same line-of-sight distance from 
the observer, but this is not necessarily the case, so some 
misidentifications can be made. \citet{MSH23} has recently shown that, 
given a long enough data set, single-dish spectra can be used to 
identify Zeeman pairs by tracking RCP and LCP peaks that drift slowly 
at the same velocity. 

Spectra of 4.7 GHz exOH masers are generally far less complicated than 
the lower energy masers from the $^2\Pi_{3/2}$ levels, often displaying 
only a single spot of emission. This makes it possible to study the 
properties of these masers, such as temporal variations, using 
single-dish telescopes. Measurements of 4.7 GHz exOH masers have often 
been observed using circular polarization feeds, but to date no Zeeman 
splitting has been identified. Linear polarization has been reported 
in only one 4.765 GHz source thus far; in the 1990s a set of flaring 
masers in Mon R2 were found to have linear polarization of $\sim 
14$\% \citep{SCH98} in two separate spots. A later flare from a 
new spot had the same percentage of linear polarization, leading 
to the suggestion that the exOH masers were amplifying background 
radiation that was polarized, rather than being intrinsic to the 
maser due to magnetic fields. 

In this paper we report the first detection of Zeeman pairs in two 
4.765 GHz exOH masers, indicating that a magnetic field permeates at 
least some part of the masing column of gas. In \S \ref{sec:sou} we 
provide brief details of the two sources in which circular polarization 
has been found, and in \S \ref{sec:obs} details of our observations 
are presented. In \S \ref{sec:res} our results are presented, including 
a discussion on skewness and how we measured it. In \S \ref{sec:con} 
our conclusions are presented.

\section{Sources} \label{sec:sou}
In a survey of 20 sources using the Green Bank Telescope (GBT) and 
C-band receiver, we found 4.765 GHz masers in eight sources, of which 
two had circular polarization. These statistics are small so should 
be treated with caution, but they do suggest that Zeeman splitting in 
these masers is not necessarily rare. However, it is weak and sensitive 
observations are required to extract this faint signal from the 
background noise. 

In a search towards IRAS sources identified as star-forming regions by 
their colours, H$_2$O masers were found in \object{IRAS 05358+3543}, 
also known as \object{G173.482+2.446} (hereafter G173), by \citet{WBH88}. 
Follow-up observations revealed the presence of weak ($S_{\nu} < 0.8$ Jy) 
lines of 100\% circularly polarized 1.665 GHz gsOH masers at velocities 
between --10 and --16 \kms. No 1.667 GHz lines were found at that time, 
but \citet{EFC07} did discover a single line with LCP at $v = -10.53$ 
\kms. At the time of these observations, the 1.665 GHz flux density peaked 
at 2.82 Jy and covered a range of velocities from --8.5 to --16.5 \kms. 
The first detection of 4.765 GHz exOH maser emission in G173 was made by 
\citet{QSB22} at velocities of $v = -16.83$ and --16.31 \kms\ with flux 
densities of 0.53 and 3.41 Jy respectively. 

The other source in which we found Zeeman splitting is \object{G213.705--12.60} 
which is also known as \object{Mon R2} IRS 3 (hereafter Mon R2). It underwent 
a major flaring episode at 4.765 GHz in the 1990s \citep{SCH98, S03}. In 
addition to regular monitoring using the 26m telescope of the Hartebeesthoek 
Radio Astronomy Observatory (HartRAO), a few target-of-opportunity (ToO) 
observations of this source were made using MERLIN. An unexpected discovery 
from those events was that the exOH masers had 14\% linear polarization 
\citep{SCH98}. No circular polarization was detected by MERLIN to a level 
below 1\% of the flux density. In 1997 December the maser flux density 
reached a peak of $\sim 80$ Jy, but only displayed linear polarization. 
This is the only 4.765 GHz maser that has previously been reported to have 
polarization. 

By the end of 1998 the exOH Mon R2 masers dropped below detectable levels 
for HartRAO and MERLIN. The 4.765 GHz maser was detected again in May 2006 
\citep{FZS06}, indicating another flaring episode. The 4.765\,GHz masers 
have been monitored at HartRAO since 2012 and have undergone several 
flaring episodes during this period. The most recent flaring episodes 
was detected while observing with the GBT in 2023 and will be reported 
separately (Fallon and Smits, in preparation), but here we report the 
first detection of circular polarization in Mon R2, which has appeared 
during the current flaring episode.

Coordinates and central velocities of the spectra for the two star-forming regions 
that have observable Stokes V signals, are listed in Table~\ref{tab:source}. 

\begin{table*}
	\centering
	\caption{Source name, J(2000) coordinates and central velocity of the spectra.}
	\label{tab:source}
	\begin{tabular}{lccc} 
		\hline
		\multicolumn{1}{c}{Source}   &R.A.(J2000)   &Dec.(J2000)   &Velocity   \\
            \multicolumn{1}{c}{ID}       &hh mm ss.s    &$\degr\ \ '\ \ ''$   &km s$^{-1}$   \\
		\hline
            G173.482+2.446 = G173     &05 39 13.0   &+35 45 51.0   &--16.7 \\
            G213.705--12.60 = Mon R2  &06 07 47.8   &--06 22 56.5  &+10.7  \\
		\hline
	\end{tabular}
\end{table*}


\section{Observations} \label{sec:obs}
All observations reported here were made with the 100m GBT using the 
C-band receiver with the Versatile GBT Astronomical Spectrometer (VEGAS) 
backend in full Stokes mode. Each observing session consisted of a 
single source together with B0529+075 as a flux density calibrator 
and 3C138 as a linear polarization calibrator. The Hi-Cal noise diode 
was used for all except three observations: G173 and Mon R2 on 6 and 21 
Jul 2023 for which the Lo-Cal noise diode was used. Subsequent 
observations were made using the Hi-Cal noise diode with an expectation 
that this would result in improved measurement stability. However, the 
ongoing observations have not indicated improvements in accuracy or 
stability using the Hi-Cal noise diode. The setup and calibrators took 
up about 50 mins, giving on-source observations of up to 1.5 hrs. All 
scans were checked individually for RF noise, and, when present, those 
scans were discarded, so most sessions ended up being shorter than 
this. Programs to analyze position- and frequency-switching 
observations to obtain full Stokes polarization were developed as part 
of project AGBT20B-424 and are described in GBT Memo 306 \citep{Fal22}. 
Calibrators were observed using position-switching by offsetting from 
the target source by $8'$, while spectra were obtained using 
frequency-switching of $-2.9$ MHz (approximately the GBT default 
value of --0.25*bandwidth) to maximise the on-source time. 

The VEGAS backend contains eight bands, each of which is operated in 
mode 15, providing a bandwidth of 11.72 MHz over 32\,768 channels, 
corresponding to a frequency resolution of 357.7 Hz. This gives a 
velocity resolution of 0.0225 \kms\ and a velocity range of $\sim 
-300$ to $+300$ \kms\ at 4.765 GHz. The rest frequency used for the 
$F = 1 \rightarrow 0$ transition in the $^2\Pi_{1/2}\ J = \frac{1}{2}$ 
level of exOH was 4.765\,562 GHz \citep{DMB77}. Two other channels 
were used to look at the 4.660 and 4.750 GHz lines but no maser 
detections were found above a $1\sigma$ level of the RMS noise listed 
in Table \ref{tab:Gauss_I} for each observation at these frequencies. 

We followed a standard procedure to analyse the spectra which were 
recorded every 2 mins. Each spectrum was first shifted and folded and 
then calibrated using the standard value. A baseline was fitted to the 
velocity range for G173 between $[-32, -22]$ and $[-11, -1]$ \kms\ 
and for Mon R2 over $[-18, +2]$ and $[+18, +30]$ \kms, using a third 
order polynomial; this broad exclusion range for Mon R2 ensured that 
the thermal component was not included in the baseline subtraction 
process. Our Mueller matrix was then applied to the individual spectra, 
which were then added together to produce the on-sky $I, Q, U, V$ 
measurements.  

We make use of the Mueller matrix for the C-band system as described 
by \citet{FSG23}, with some modifications. Firstly, the Mueller 
matrix component $M_{4,4}$ is set to --1 rather than +1 to give the 
polarization orientation in agreement with other instruments. Because 
the circular polarization of calibrators 3C138 and 3C286 is taken to 
be zero, the orientation of Stokes $V$ was not set during the Mueller 
matrix calibration. The --1 setting aligns with the --1 of the 
$M_{2,2}$ and $M_{3,3}$ components, and the GBT receiver's orientation. 
Due to instability in both the intensity and polarization between 
the single position-switched scans of 3C138, in particular in the latter 
part of 2024 and during 2025, several 3C138 observations were excluded 
and an average Mueller matrix value was used initially for all observations. 
Further checks on the Mueller matrix values were made with observations of 
3C286 in April and May 2025. A further small Mueller matrix refinement was 
made when fitting $dI/dv$ to Stokes $V$ (Method 3 in section \ref{sec:delta}). 
The fitting process highlights the need for Mueller matrix corrections, 
described by \citet{TH82,E98}, which is due to leakage of Stokes $I$ into 
Stokes $V$. Once this Mueller matrix adjustment was implemented, all 
analyses were re-run.

All Gaussians fitted to our data made use of the least squares fitting 
routine in GBTIDL.

\section{Results} \label{sec:res}

\subsection{G173.482+2.446 = G173}
\begin{figure}[ht!]
\centering
\scalebox{0.7}{\plotone{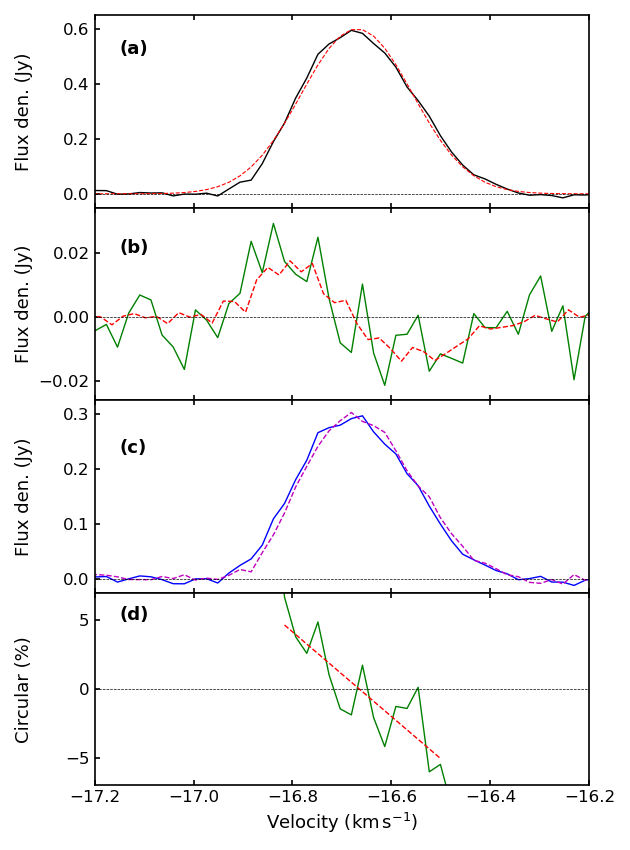}}
\caption{Spectra of (a) Stokes $I$, (b) Stokes $V$, (c) RCP and LCP, 
and (d) $V/I$ for the 4.765 GHz exOH maser in G173.} 
\label{fig:G173}
\end{figure}

Only one observation was made of this source, the date and duration of 
which are presented in Table~\ref{tab:Gauss_I} labelled G. Our 4.765 GHz 
$I$ spectrum is shown in black in Fig.~\ref{fig:G173}(a). Parameters of 
a Gaussian profile fitted to the Stokes $I$ component, shown in red on 
the plot, are listed in Table~\ref{tab:Gauss_I}. This maser, with a flux 
density of only \added{$S_{\nu} = 0.6$} Jy, is weak and, hence, has a 
poor signal-to-noise ratio (SNR). Our measured flux densities and 
velocities of our fitted Gaussians are different to those of 0.53 
and 3.41 Jy at $v = -16.83$ and --16.31 \kms\ reported by \citet{QSB22}, 
so there is clearly some variability in the source between the two sets 
of observations. 

No linear polarization was found above the noise level, but we did 
find circular polarization at a level of 5\% as can be seen from 
the green line in Fig.~\ref{fig:G173}(b) which is noisy due to the low
SNR for this data. This $V$ spectrum shows the characteristic sine-wave 
shape expected when the velocity shift between RCP and LCP Gaussian 
profiles is small, as first pointed out by \citet{GSS60} (see their 
Fig.~1). The red line is a plot of scaled $dI/dv$ determined from the 
Stokes $I$ profile in Fig.~\ref{fig:G173}(a) and is a good fit to the 
green line. \added{In Fig.~\ref{fig:G173}(c) the solid blue line is the 
RCP spectrum and the dotted line is the LCP spectrum.} The green line 
in Fig.~\ref{fig:G173}(d) is a plot of Stokes $V/I$ as a percentage, 
and the red line is a straight line fit to this data over the width of 
the emission.

The sine-shaped curve of the $V$ spectrum and the straight-line fit 
to the $V/I$ plot \added{are consistent with (but not exclusive to)} 
a single spot of circularly polarized emission (see Appendix A). 
However, the sine-shaped curve lacks the symmetry of a pure theoretical 
treatment. This asymmetry will be addressed later when we discuss the 
skewness parameter presented in Table~\ref{tab:Gauss_I}. 

The circular components of the maser signal were determined using RCP = 
$(I + V)/2$ and LCP = $(I - V)/2$. The parameters of single Gaussians 
fitted to these profiles are listed in Table~\ref{tab:RCP_LCP} in the 
top row labelled G. Following standard practice, the Full Width Half 
Maximum (FWHM) of the maser profiles are listed rather than the 
standard deviation $\sigma$ of the Gaussians. They are related by
FWHM $= \sqrt{2\ln2}\sigma = 2.355\sigma$. 

\subsection{G213.705--12.60 = Mon R2}
\begin{figure}[h!]
\centering
\scalebox{0.7}{\plotone{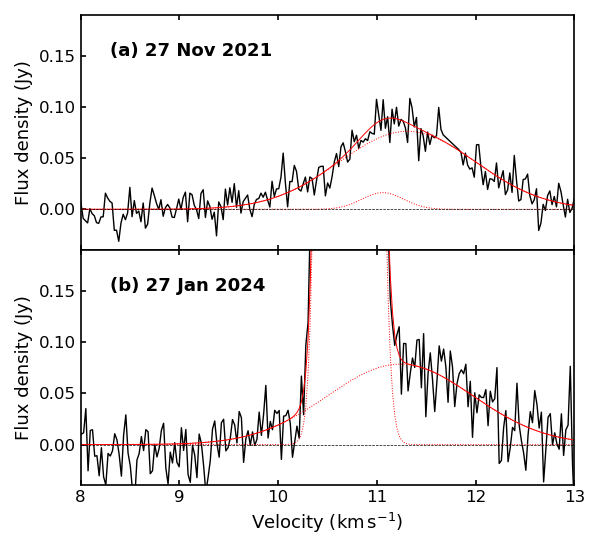}}                {}
\caption{The thermal emission in the 4.765 GHz Mon R2 spectrum on a) 27 
November 2021 and (b) 27 January 2024. \added{The black line is the 
observed data, the dotted red lines are fits to the individual Gaussians, 
and the solid red line is a sum of the individual components.}}
\label{fig:thermal}
\end{figure}

A new feature in the $I$ spectra of the 4.765 GHz exOH emission in Mon 
R2 is the presence of a broad thermal component. This profile has been 
observed in all our spectra taken with the GBT (including one where 
there was no significant maser emission). In Fig.~\ref{fig:thermal} 
this thermal component is shown on 27 Nov 2021 before the recent flaring 
episode began, and on 27 Jan 2024 when the maser had a flux density 
\added{$S_{\nu} = 19.11$} Jy. This signal is too weak to be identified in the 
HartRAO spectra, and because thermal emission usually occurs over a large 
area, it would have been resolved out by MERLIN observations. Hence, if 
this thermal component has been present since the 1990s, it would not 
have been noticed in previous observations. Because the $Q,\ U$ and $V$ 
spectra are the difference between two orthogonal components, the 
thermal component, which is assumed to be unpolarized, is automatically 
absent. 

When the maser started flaring again in mid 2023, the thermal profile 
was still present and had to be subtracted from the spectra to determine 
the pure maser emission. To achieve this, the $I$ spectrum was fitted 
with two or three Gaussians: one for the thermal profile and the other 
one or two to model the maser emission. For all observations except that 
on 2023 Jul 06, a single Gaussian fitted to the maser emission left 
large residuals, but when two Gaussians were fitted the residuals 
consisted only of noise. This fitting procedure gave the parameters for 
the thermal Gaussian profile which could then be subtracted from the $I$ 
spectra to leave the maser components.

In Table~\ref{tab:Gauss_I} the observation dates of Mon R2 are listed 
together with the time spent on-source, and the ensuing RMS noise in 
our spectra. Gaussian parameters fitted to the thermal and Stokes $I$ 
components for each session are also listed, together with the skewness 
of the $I$ profile which will be discussed in more detail in \S 
\ref{sec:skewness}.  

The thermal emission variability between the observations is larger 
than the uncertainties in the fitted Gaussian parameters. This could 
be due to extracting a weak signal from the maser signal which is two 
orders of magnitude larger than the thermal flux density. An 
observation made on 2021 Nov 27 had almost no maser emission so the 
spectrum was dominated by the thermal component. A Gaussian profile 
fitted to this spectrum had a flux density \added{$S_{\nu} = 76(6)$ 
mJy}, a central velocity $v = 11.32(4)$ \kms, and a FWHM $\Delta v = 
1.60(8)$ \kms, where the figures in parentheses are the uncertainties 
of the least significant figures. Adding the thermal components of 
all our 18 observations of Mon R2 together and then fitting a Gaussian 
gives parameters of peak flux density \added{$S_{\nu} = 87(1)$} mJy, 
velocity of peak $v = 11.143(12)$ \kms\ and FWHM $\Delta v = 1.71(3)$ 
\kms. This is noticeably different to the values from 2021 Nov 27, 
which could indicate that the maser has an influence on the thermal 
component. This will require further investigation.
   
\begin{table*}[h]
	\centering
	\caption{Details of observations and Gaussian fits to G173 (G) and 
         Mon R2 (1 -- 16). The first column labels the observations to 
         identify the entry in other tables. There is no discernible thermal 
         component in the G173 spectrum.}
	\label{tab:Gauss_I}
	\begin{tabular}{|c|c|cc|ccc|ccc|c|} 
		\hline
\#  &Date         &Obs.        &      
  &\multicolumn{3}{|c|}{Gaussian fit to Thermal Comp} 
  &\multicolumn{3}{c|}{Single Gaussian fit to Stokes $I$} &2 Gaussian\\
    &Observed   &Time    &RMS   &\multicolumn{1}{c}{Flux density}   &Velocity   &FWHM   
         &Flux density  &Velocity   &FWHM  &fit to $I$    \\
    &\multicolumn{1}{|c|}{yyyy mmm dd}  &min   &mJy  &\multicolumn{1}{c}{Jy}   &\kms    &\kms   
         &Jy  &\kms  &\kms  &Skewness    \\
		\hline
G   &2023 Apr 23   &68   &7 &            &           &         &0.600(3)   &--16.6697(7)   &0.2645(17)   &0.17 \\ \hline
1     &2023 Jul 06     &28     &42     &0.099(10)     &11.14(8)     &1.62(16)     &12.77(2)     &10.7216(2)     &0.2999(5)     &-0.001\\
2     &2023 Jul 21     &50     &30     &0.103(7)     &11.11(5)     &1.67(10)     &11.38(1)     &10.7201(2)     &0.2993(4)     &0.015\\
3     &2023 Dec 23     &70     &27     &0.087(6)     &11.19(6)     &1.57(10)     &16.27(2)     &10.7017(2)     &0.2854(5)     &0.146\\
4     &2024 Jan 27     &90     &22     &0.079(5)     &11.24(6)     &1.73(11)     &19.11(3)     &10.6986(2)     &0.2822(5)     &0.166\\
5     &2024 Feb 20     &54     &35     &0.096(9)     &11.13(7)     &1.68(13)     &20.72(3)     &10.6973(2)     &0.2797(5)     &0.180\\
6     &2024 Apr 03     &70     &29     &0.082(7)     &11.08(6)     &1.82(13)     &21.10(3)     &10.6943(2)     &0.2764(5)     &0.194\\
7     &2024 Apr 28     &48     &32     &0.083(7)     &11.34(8)     &1.50(15)     &25.49(4)     &10.6906(2)     &0.2744(5)     &0.216\\
8     &2024 May 22     &68     &27     &0.084(6)     &11.10(6)     &1.82(12)     &21.78(4)     &10.6886(2)     &0.2715(5)     &0.238\\
9     &2024 Aug 18     &42     &46     &0.098(12)     &11.11(9)     &1.72(17)     &21.85(4)     &10.6863(3)     &0.2628(6)     &0.276\\
10     &2024 Oct 05     &48     &42     &0.074(8)     &11.25(11)     &2.11(21)     &18.85(4)     &10.6914(3)     &0.2619(6)     &0.273\\
11     &2024 Nov 05     &38     &35     &0.107(10)     &11.08(6)     &1.61(11)     &17.08(3)     &10.6950(2)     &0.2612(6)     &0.248\\
12     &2025 Jan 27     &66     &27     &0.086(6)     &11.11(5)     &1.70(11)     &15.17(2)     &10.7073(2)     &0.2663(5)     &0.203\\
13     &2025 Feb 24     &38     &42     &0.115(13)     &11.00(6)     &1.40(12)     &17.12(3)     &10.7121(2)     &0.2679(5)     &0.182\\
14     &2025 Mar 19     &70     &26     &0.078(6)     &11.06(6)     &1.98(13)     &20.28(3)     &10.7149(2)     &0.2712(5)     &0.173\\
15     &2025 Apr 11     &64     &36     &0.083(9)     &11.21(8)     &1.52(15)     &22.98(3)     &10.7171(2)     &0.2747(5)     &0.148\\
16     &2025 May 02     &60     &28     &0.100(7)     &11.05(5)     &1.68(10)     &26.21(4)     &10.7187(2)     &0.2773(5)     &0.145\\
\hline
	\end{tabular}
\end{table*}

\begin{figure}[ht!]
\plottwo{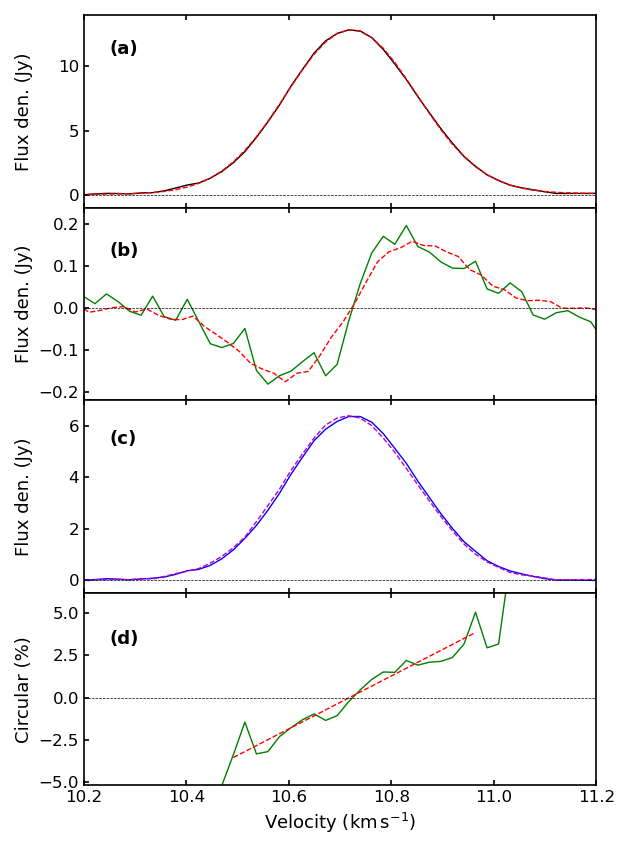}{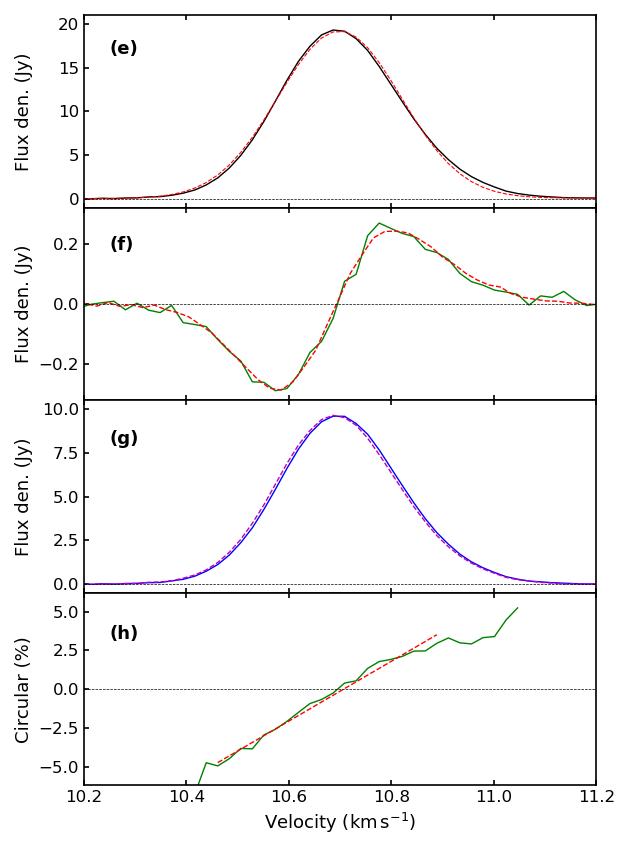}
\caption{Spectra of (a) \& (e) Stokes $I$, (b) \& (f) Stokes $V$, 
         (c) \& (g) RCP and LCP, and (d) \& (h) $V/I$ of the 4.765 GHz exOH 
          maser in Mon R2 on 06 Jul 2023 and 27 Jan 2024.}
\label{fig:MonR2}
\end{figure}

Plots of the Stokes $I$, Stokes $V$, RCP and LCP, and $V/I$ for two 
epochs are shown in Fig.~\ref{fig:MonR2}. The black $I$ component is 
the data, fitted with a single Gaussian in red. The green $V$ profile 
displays the sine-wave shape expected for the difference between two 
closely spaced Gaussians, and is fitted well by the derived scaled 
red $dI/dv$ curve. In the 2023 Jul 06 spectrum, the $V$ shape is 
reasonably symmetrical while in all other Mon R2 (and G173) observations 
reported here, there is an asymmetry between the positive and negative 
sections of the $V$ profile. In the bottom plot the green $V/I$ data is 
fitted with a red straight line. 

The circular components of the maser signal were determined using 
RCP = $(I + V)/2$ and LCP = $(I - V)/2$. The parameters of single 
Gaussian profile fitted to these data are presented in 
Table~\ref{tab:RCP_LCP}. The velocity separation $2\delta$ between 
the RCP and LCP components is small; $\delta$ measured from the 
Gaussian fits is listed in Table~\ref{tab:delta} together with 
values determined by two other methods, which are described in \S 
\ref{sec:delta}.

\begin{table*}
	\centering
	\caption{Gaussian parameters of RCP \& LCP components for G173 (G) 
             and Mon R2 (1 -- 16).}
	\label{tab:RCP_LCP}
	\begin{tabular}{|c|ccc|c|ccc|c|c|} 
		\hline		 
\#   &\multicolumn{3}{|c|}{Single Gaussian fit to RCP}	&2 Gaussian &\multicolumn{3}{c|}
{Single Gaussian fit to LCP}  &2 Gaussian		&Amplitude	\\		
  &Flux density &Velocity  &FWHM  &fit to RCP  &Flux density &Velocity  &FWHM  &fit to LCP &ratio\\
  &Jy    &\kms      &\kms  &Skewness  &Jy   &\kms      &\kms  &Skewness    &$f$\\
\hline
G   &0.299(2)   &--16.6743(9)  &0.267(2)   &0.16 &0.302(2)   &--16.6653(9)  &0.262(2)  &0.19   &1.009(11)\\ \hline
1     &6.39(1)     &10.7243(3)     &0.2998(6)     &-0.001     &6.39(1)     &10.7189(2)     &0.2999(5)     &0.000     &1.000(2)\\
2     &5.69(1)     &10.7227(2)     &0.2992(5)     &0.015     &5.69(1)     &10.7174(2)     &0.2992(4)     &0.013     &1.000(2)\\
3     &8.14(1)     &10.7043(2)     &0.2854(5)     &0.144     &8.14(1)     &10.6991(2)     &0.2854(5)     &0.149     &1.000(2)\\
4     &9.56(1)     &10.7013(2)     &0.2819(5)     &0.169     &9.55(1)     &10.6959(2)     &0.2822(5)     &0.164     &0.999(2)\\
5     &10.35(2)     &10.6998(2)     &0.2802(5)     &0.176     &10.37(2)     &10.6948(2)     &0.2791(5)     &0.185     &1.002(2)\\
6     &10.56(2)     &10.6970(2)     &0.2762(5)     &0.190     &10.55(2)     &10.6916(2)     &0.2764(5)     &0.199     &1.000(2)\\
7     &12.75(2)     &10.6932(2)     &0.2742(5)     &0.203     &12.74(2)     &10.6879(2)     &0.2744(5)     &0.230     &1.000(2)\\
8     &10.89(2)     &10.6913(2)     &0.2717(5)     &0.231     &10.90(2)     &10.6860(2)     &0.2712(5)     &0.243     &1.001(2)\\
9     &10.93(2)     &10.6891(3)     &0.2627(6)     &0.267     &10.93(3)     &10.6835(3)     &0.2628(7)     &0.281     &1.000(3)\\
10     &9.43(2)     &10.6944(3)     &0.2619(6)     &0.266     &9.43(2)     &10.6885(3)     &0.2617(6)     &0.275     &1.000(3)\\
11     &8.55(2)     &10.6977(3)     &0.2607(6)     &0.241     &8.54(2)     &10.6922(3)     &0.2615(6)     &0.254     &0.999(3)\\
12     &7.59(1)     &10.7102(2)     &0.2660(5)     &0.210     &7.58(1)     &10.7044(2)     &0.2664(5)     &0.193     &0.999(2)\\
13     &8.56(2)     &10.7147(2)     &0.2678(6)     &0.183     &8.56(2)     &10.7095(3)     &0.2679(6)     &0.182     &1.000(3)\\
14     &10.14(2)     &10.7179(2)     &0.2711(5)     &0.170     &10.15(2)     &10.7119(2)     &0.2710(5)     &0.172     &1.000(2)\\
15     &11.49(2)     &10.7200(2)     &0.2745(5)     &0.149     &11.49(2)     &10.7142(2)     &0.2747(5)     &0.147     &1.000(2)\\
16     &13.11(2)     &10.7215(2)     &0.2770(5)     &0.146     &13.10(2)     &10.7159(2)     &0.2774(5)     &0.145     &0.999(2)\\
    \hline
	\end{tabular}
\end{table*}

\subsection{Temporal changes in the Mon R2 line profile} 
\label{sec:SkewChange}
\begin{figure}[ht!]
\centering
\scalebox{0.7}{\plotone{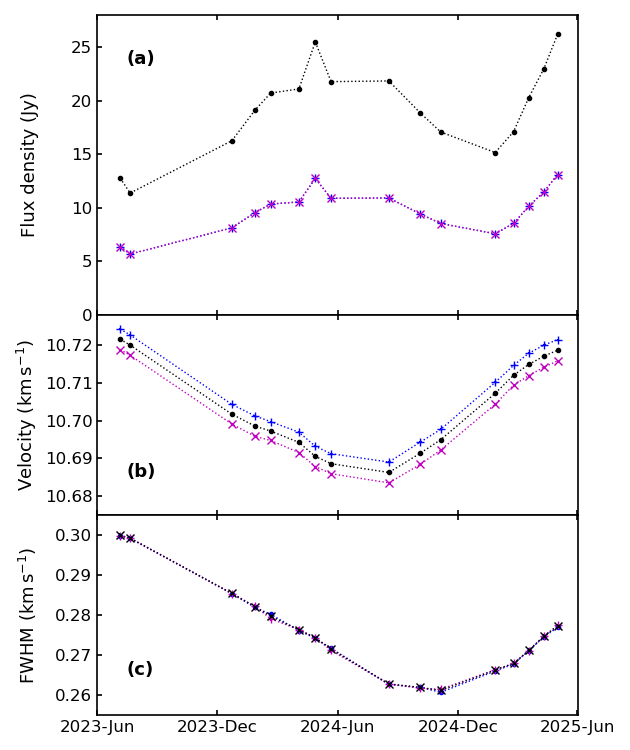}}
\caption{Time variation of (a) the peak flux density, (b) peak 
velocity $\mu$, and (c) FWHM for the Mon R2 Stokes $I$, RCP and 
LCP maser profiles. The error bars are included, but are smaller 
than the markers.} 
\label{fig:FVW}
\end{figure}

The evolution of the Mon R2 flaring can be traced using the data 
from Tables \ref{tab:Gauss_I} and \ref{tab:RCP_LCP}. The peak flux 
density $I$, the peak velocity $\mu$ and the FWHM are plotted in 
Fig.~\ref{fig:FVW}(a), (b), and (c), respectively, together with 
the values for the RCP and LCP components. The amplitudes of the 
circular components are similar and follow the behaviour of the 
Stokes $I$ flux density. The peak velocities $\mu$ of the 
three components all vary synchronously with a fairly constant 
separation between each of them. The FWHM of the single Gaussain 
profiles show that all three components vary synchronously. This 
gives us confidence that the fitted functions are reliable measures 
of the line profiles.

The variations in the flux density are an intrinsic property 
of the maser, whereas the variations in the peak velocity $\mu$ and 
the FWHM appear to be related to the skewness. Long term 
monitoring of masers show that the central velocity may drift slowly 
with time \citep{MSH23}, but not usually over the short time periods 
of our data. The width of maser profiles can vary if they are made up 
of multiple spots of emission that cannot be resolved in space or 
velocity, but the sine-wave shape of the $V$ curve is consistent with 
a single spot of maser emission which should remain constant. A more 
likely explanation for the temporal changes in $\mu$ and FWHM is that 
they are due to the influence of the skewness which affects the values 
of these two fitted Gaussian parameters. Skewness is discussed in the 
next section.

\subsection{Maser line profile skewness} \label{sec:skewness}
The $f$ parameter listed in Table \ref{tab:RCP_LCP} is a measure of 
the difference in amplitude between the RCP and LCP profiles. Because 
the $f$ values all have values close to unity, the asymmetry between 
the positive and negative portions of the Stokes $V$ sine curve 
cannot be due to amplitude differences between the circular components. 
An unexpected observation from this study is that the Stokes $I$, RCP 
and LCP maser line profiles are distorted Gaussians. The models 
of interstellar H$_2$O masers developed by \citet{WSS02} examined the 
second order deviation of Gaussian symmetry known as kurtosis, but did
not look at the third order asymmetry known as skewness. Unsaturated 
methanol lines have been found to display kurtosis \citep{MMW03}. Both 
the G173 and Mon R2 sources have a single maser line profile and show 
a measure of skewness that is evident when trying to fit the spectra 
with a single Gaussian. 

Skewness in Gaussian profiles of maser lines has been considered 
previously \citep{VvL05}, however, for comparison of masers with 
different intensity and FWHM, it is important to use a normalized 
skewness formula. The asymmetry of the maser lines was measured 
using Fisher's moment coefficient of skewness \citep{JG98}, given by
\be \label{eq:skewness}
\mbox{Skewness} = \frac{\frac{1}{\Sigma^n_{i=1} S_i}\Sigma^n_{i=1} S_i \left(
\frac{v_i - \bar{v}}{\sigma_s} \right)^3}{\left[\frac{1}{\Sigma^n_{i=1} S_i}
\Sigma^n_{i=1} S_i \left(\frac{v_i - \bar{v}}{\sigma_s}\right)^2\right]^{3/2}}
= \frac{\frac{1}{\Sigma^n_{i=1} S_i}\Sigma^n_{i=1} y_i \left(v_i - \bar{v} 
\right)^3}{\left[\frac{1}{\Sigma^n_{i=1} S_i}\Sigma^n_{i=1} S_i \left(v_i - 
\bar{v}\right)^2\right]^{3/2}},
\ee
where $\bar{v}$ is the velocity distribution mean and $\sigma_s$ is 
its standard deviation. The spectrum data are represented by velocity 
$v_i$ and intensity $S_i$ in each channel. Because noise either side 
of the maser line contributes to the value of skewness, the calculation 
is sensitive to $\bar{v}$ and the velocity range used. After testing, 
a range of $\bar{v} \pm 5\sigma_s$ was used. This is sufficient to 
include all the skewness of the distribution, whilst minimizing the 
noise on either side of the maser line. The center velocity $\mu$ and 
standard deviation $\sigma$ of a Gaussian fit to the maser line are 
not the same as $\bar{v}$ and $\sigma_s$ of the skewed distribution. 
Hence, an iterative process was used to determine values for $\bar{v}$ 
and $\sigma_s$, using the Gaussian fit parameters as starting values. 

\begin{figure}[ht!]
\centering
\scalebox{0.7}{\plotone{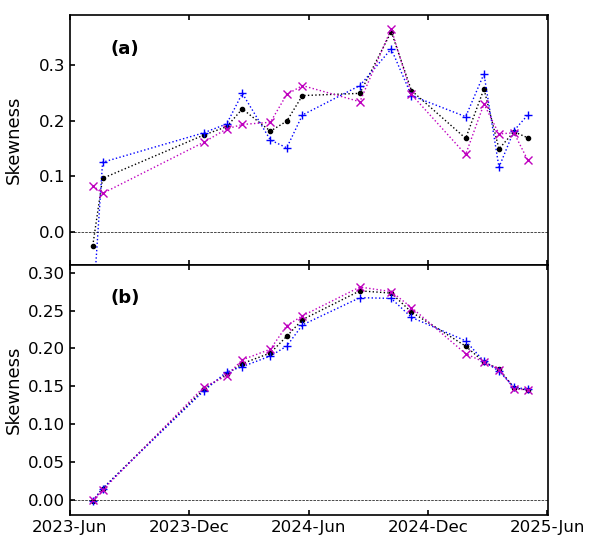}}
\caption{Mon R2 maser line profile skewness generated from (a) the 
spectral data and (b) a two-Gaussian fit to the skewed profiles.} 
\label{fig:skew}
\end{figure}

The skewness of the Mon R2 Stokes $I$, RCP and LCP Gaussian profiles 
determined using Eq.~(\ref{eq:skewness}) are plotted in 
Fig.~\ref{fig:skew}(a). Noise in the data, in particular at the tails 
of the distribution, affects the skewness calculation. This 
introduces deviations in the calculated skewness which produces 
erratic behaviour, as can be seen in the plot. Overall, the skewness 
in all three components follow a similar pattern, showing a slow rise 
followed by a fall. 

To try and improve the calculations, an alternative method of measuring 
skewness was tested. In all cases, we found we could fit the skewed 
profiles with two Gaussians, leaving residuals no larger than the noise. 
It is beyond the scope of this paper to determine if the two Gaussian 
fit is a general feature of a skewed Gaussian profile, or if we were 
just fortunate to be able to restrict the fits to only two components. 
These two Gaussian were added together to produce a profile that is 
not affected by the noise associated with the observed data. This 
summed profile was then analysed using Eq.~(\ref{eq:skewness}). 
The skewness calculated using the two-Gaussian model is shown in 
Fig.~\ref{fig:skew}(b). Comparision with Fig.~\ref{fig:skew}(a) shows 
a much smoother change in skewness. 

The skewness plotted in Fig.~\ref{fig:skew}(b) has an inverse 
behavior to that of the center velocity $\mu$ and FWHM  shown in 
Figs~\ref{fig:FVW}(b) and (c), suggesting that the changes in
these parameters are related to variations in the skewness of the
profile. Using our two-Gaussian model, the peak velocity 
of the combined spectra was calculated, and it also underwent 
shifts matching the behaviour of the single Gaussian fits. Therefore, 
it appears that the skewness is affecting the physical conditions 
in the column of masing gas. It is noteworthy that the skewness is 
not related to the intensity of the maser, but it does affect the 
peak velocity and FWHM. This is clearly an area that needs to be 
investigated further.

Our analysis suggests several reasons why our data represent skewed 
profiles as opposed to two overlapping maser lines.

Modeling Stokes $V$ using each profile from the two Gaussian 
model and then adding them together, produces $V$ spectra that 
do not match our observed data.

The smooth change in skewness seen in Mon R2 (see Fig.~\ref{fig:skew}b), 
and the corresponding changes in the maser velocity and FWHM, appear 
to be related. In contrast, heights for the two fitted Gaussians display  
a less coherent variation over time.

Finally, e-MERLIN imaging of the Mon R2 maser on 2024 Oct 14 
produced only one spot of emission at its resolution of 40 mas. 
Assuming a distance of 830 pc to Mon R2, this corresponds to $\sim 
33$ AU. This does not rule out the possibility of two unresolved  
maser spots, but a single spot is consistent with our results.

\subsection{Determination of velocity between circular components}
\label{sec:delta}
The separation in velocity between the circular components is equal 
to $2\delta$ where $\pm \delta$ is the shift in velocity due to the 
Zeeman effect on the RCP and LCP components from the zero $B$ field 
position. The value of $\delta$ has been determined using three 
different methods, not all of which are independent. However, the 
different methods do provide a consistency check on the various 
values. 

\begin{description}
\item[Method 1] The RCP and LCP profiles have been fitted with 
single Gaussians to determine the amplitude $A$, central velocity 
$\mu$ and FWHM  of the lines. Our velocity spectral resolution of 
0.0225 \kms\ facilitated using about 30 channels to fit the line 
profiles, thereby producing velocity uncertainties lower than the 
resolution of the spectra. 

\item[Method 2] In Appendix \ref{appendix:slope} it is shown that 
if the RCP and LCP profiles are formed from single spots of emission 
with Gaussian shapes, then, provided $\delta \ll \sigma$, $V/I$ is 
a straight line with a slope $m = \delta/\sigma^2$ 
( Eq.~(\ref{eq:V/I})). By fitting a straight line to the observed 
$V/I$ plot, and using $\sigma$ from the Gaussian fitting procedure, 
a value for $\delta$ can be determined. Because this method relies 
on results from the Gaussian fitting procedure it is not an 
independent method. Unfortunately, skewness has an influence on the 
straight line and hence the method's accuracy which we have not 
found a way to cater for.

\item[Method 3] When the separation between RCP and LCP is small, 
the derivative of the Stokes $I$ spectrum with respect to $v$ is 
proportional to the Stokes $V$ spectrum. Specifically for Gaussians 
of equal amplitude $dI/v = -V/\delta$, as shown in Appendix 
\ref{appendix:derivative}. As seen in Figs \ref{fig:G173} and 
\ref{fig:MonR2}, the scaled $dI/dv$ fits the $V$ spectrum remarkably 
well. This approach has been used previously to determine magnetic 
field strengths of Zeeman split lines \citep{TH82,KC86,CTG93,YRG96}. 
\end{description}

\begin{figure}
\centering
\scalebox{0.7}{\plotone{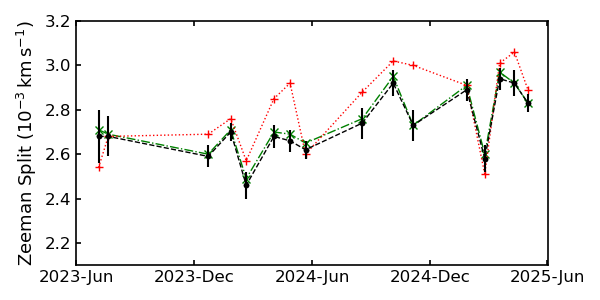}}                {}
\caption{Temporal variations of the velocity shifts due to the Zeeman 
         effect in Mon R2. Method 1 is given by green $\times$ and a 
         dash-dotted line, Method 2 is given by red $+$ and a dotted 
         line, and Method 3 by black $\bullet$ and a dashed line. The 
         error bars are from Method 3 in Table \ref{tab:delta}.}
\label{fig:Zsplit}
\end{figure}

The values of $\delta$ determined using the above three methods 
produce similar results as seen in Table \ref{tab:delta} for G173 (G) 
and Mon R2 (1 -- 16) and plotted in Fig.~\ref{fig:Zsplit}. Method 1 
can be applied under all circumstances, whereas Method 2 and 3 are 
only applicable if $\delta \ll \sigma$, which is the situation for 
these measurements. Method 1 and Method 3 have similar values but 
the uncertainties for Method 1 are larger than for Method 3. Because 
the profiles are skewed Gaussians, the straight line relationship 
for $V/I$ is affected and produces values that differ from the other 
two methods, but not by much. Method 3 relies on fitting only one 
parameter, thereby producing the smallest uncertainties. The strength 
of the line-of-sight (LOS) magnetic field $B_{\textrm LOS}$ has been 
calculated from the values of Method 3. 

\begin{table*}[ht!]
	\centering
	\caption{Velocity shift $\delta$ for G173 (G) and Mon R2 (1 -- 16) 
    using three different methods as discussed in the text. The magnetic 
    field $B_{LOS}$ is calculated from Method 3 using a conversion factor 
    0.0315 \kms\ G$^{-1}$.}
	\label{tab:delta}
	\begin{tabular}{|c|ccc|c|} 
		\hline		 	
\#  &\multicolumn{3}{c|}{$\delta \ [10^{-3}$ \kms ]} &$B_{\textrm LOS}$\\
\  &Method 1       &Method 2    &Method 3     &mG\\
\hline
G     &--4.5(18)     &--3.9(7)     &--4.3(5)     &--137(15) \\ \hline
1     &2.71(17)     &2.54(16)    &2.68(12)    &85(4)\\
2     &2.69(13)     &2.68(9)     &2.68(9)     &85(3)\\
3     &2.60(14)     &2.69(11)    &2.59(5)     &82.2(17)\\
4     &2.71(14)     &2.76(8)     &2.70(4)     &85.9(11)\\
5     &2.49(15)     &2.57(7)     &2.46(6)     &78.0(19)\\
6     &2.70(15)     &2.85(9)     &2.68(5)     &85.3(16)\\
7     &2.69(15)     &2.92(7)     &2.66(5)     &84.7(15)\\
8     &2.65(16)     &2.60(8)     &2.62(4)     &83.3(13)\\
9     &2.76(20)     &2.88(9)     &2.74(7)     &87(2)\\
10    &2.95(18)     &3.02(10)    &2.92(6)     &92.9(19)\\
11    &2.73(18)     &3.00(11)    &2.73(7)     &87(2)\\
12    &2.91(16)     &2.91(10)    &2.89(5)     &92.0(15)\\
13    &2.60(17)     &2.51(14)    &2.58(6)     &82.2(19)\\
14    &2.97(15)     &3.01(9)     &2.94(5)     &93.6(17)\\
15    &2.92(15)     &3.06(11)    &2.92(6)     &92.7(17)\\
16    &2.83(14)     &2.89(11)    &2.83(4)     &89.8(11)\\
 \hline
\end{tabular}
\end{table*}

In Table 1 of \citet{D74}, an approximate value for the conversion 
factor $\nu/B \sim 10^{-3}$ MHz G$^{-1}$ for the separation between
the RCP and LCP components is given. For our determinations of the 
magnetic field strength $B$, which are based on $\delta$ rather than 
$2\delta$, we set $\nu/B = 5 \times 10^{-4}$ which gives a value $v/B 
= 3.15 \times 10^{-2}$ \kms\ G$^{-1}$ with which to calculate $B$ 
from our velocity measurements. This factor is still uncertain and 
could be higher or lower than this, but it is easy to modify our 
values if, or when, a more reliable value becomes available. Based 
on this conversion factor the Land\'{e} factor for the $^2\Pi_{1/2}\ 
J= 1/2$ level is $g_J = 3.5 \times 10^{-4}$ which is smaller than the 
uncertainties in the current values for $g_J$ in the gs and first 
excited state of OH.

The observed maser circular polarization results from the portion 
of the magnetic field $B$ along the observer's LOS, which for $B$ at 
an angle $\theta$ to the LOS becomes $B\cos\theta$. The value of
$\theta$ cannot be determined from the circular polarization
measurements alone, the linear component is needed for this. \added{The 
negative $B_\textrm{LOS}$ value for G173 and the positive value for 
Mon R2 indicate one field pointing away from the observer and the 
other towards the observer, depending on the convention adopted (see 
\citet{RH21} \S6.3). }

As shown in Fig.~\ref{fig:Zsplit} and Table \ref{tab:delta}, the
velocity shift values change over time, with uncertainties smaller 
than the variations. We believe these changes are real and are related 
to variations seen in the maser's linear polarization \citep{FS24}. 
These polarization changes and the mechanism causing them will be 
presented separately (Fallon and Smits, in preparation).

There are no 6.0 GHz exOH maser observations towards either of these 
sources with which to compare magnetic field measurements, but there 
are some gsOH surveys that have been done, as well as some methanol 
observations.

The VLA survey of \citet{ARM00} found four spots of 1665 gsOH masers 
towards G173, all with LCP only. The velocities of these spots differ
by 4 -- 7 \kms\ from the 4.765 GHz maser so they are probably in a 
separate position. \citet{EFC07} found one spot each of 1.665 and 
1.667 GHz lines at LCP only at velocities of $v = -10.88$ and $-10.53$ 
\kms\ which are different to those of the earlier VLA discoveries. 
Full Stokes observations were made by \citet{SG09} who found 1.665 
and 1.667 GHz masers with both RCP and LCP, but the LCP components 
were much stronger than the RCP spots, so that Stokes $V$ was nearly 
100\% LCP. No Zeeman pairs were identified. These spots did not have 
a high degree of linear polarization. VLA observations made by 
\citet{BVK21} found five spots of 1.665 GHz RCP and two of LCP but no 
obvious Zeeman pairs and the velocities were different to our 4.765 
GHz maser. At 1.667 GHz there was one spot each of RCP and LCP but 
they were at different positions so not a Zeeman pair. The velocities
of the gsOH lines have varied in each of the above observations, 
indicating that this is a source that is undergoing changes. The 
isolated RCP and LCP spots are consistent with a magnetic field 
that is stronger than 10 mG, allowing only one mode to grow.

Observations of Zeeman splitting in the methanol 6.7 GHz line were 
made using the Effelsberg 100m telescope by \citet{V08}. The 
non-paramagnetic methanol molecule has a small Land\'{e} g-factor, 
producing a separation of 0.049 \kms\ G$^{-1}$. A magnetic field
of $B = 19(2)$ mG was measured at a velocity around $v = -13$ \kms.
This is about seven times smaller than our estimate but it is also
at a different velocity and in the opposite direction to the magnetic
field we measured. 

Mon R2 was found to have 1.665 and 1.667 GHz maser spots in 
both RCP and LCP in the VLA survey of \citet{ARM00}, but none of 
these spots manifested as Zeeman pairs. The \citet{SG09} full Stokes 
survey found RCP and LCP spots in 1.665 and 1.667 GHz, dominated by 
the RCP components. The 1.667 GHz Stokes $V$ component has a sine-wave 
shape which could indicate Zeeman splitting but the position of the spots 
were not determined with enough resolution to locate them at the 
same position. The peaks of the sine wave are separated by $\sim 1$ 
\kms\ so if they were a Zeeman pair the magnetic field would be $< 
10$ mG and the maser spot velocities are slightly different to that 
of the 4.765 GHz maser, so probably not in the same region. 

There are methanol 6.7 GHz masers in the Mon R2 region, but 
they are variable and at slightly different velocities to the 4.7 GHz 
exOH maser. \citet{SVL15} detected some weak linear polarization 
($P_l < 5\%)$, and one spot with circular polarization of 0.6\%. 
The magnetic field associated with this spot was very weak.

Conventional wisdom posits that magnetic fields in star-forming regions 
lie in the range of $1 - 10$ mG. These measurements have been made 
predominantly using the Zeeman splitting of 1.7 and 6.0 GHz masers which 
are sensitive to fields in this range. The problem with this method is 
that it is not sensitive to fields much stronger than 10 mG. As the 
field strength grows, so too does the separation between the RCP and 
LCP components. If this separation becomes large enough, velocity 
coherence along the length of the maser will only occur in one component, 
the other one will not be amplified. In array maps of star-forming 
regions, isolated RCP and LCP spots are seen, but these can not be 
used to determine magnetic field strengths. This is a limitation of the 
currently-used OH masers to measure magnetic fields: they can only 
measure fields within a limited range of strength, above that only one 
component appears as a maser, and so the field strength cannot be 
determined. Stronger fields could exist, but a different tool is needed 
to measure them. Zeeman splitting of 4.7 GHz exOH masers is a tool that 
can measure stronger fields than those of the gs and 6.0 GHz exOH maser 
lines. 

The 13.441 GHz $F = 4 \rightarrow 4$ transition of the $^2\Pi_{3/2}, 
J = 7/2$ state of OH has also been used by \citet{C04} to measure magnetic 
fields in star-forming regions. These masers are rare which suggests 
they require very specific conditions for their existence. The Land\'{e} 
g-factor for this transition is smaller than for the 1.7 and 6.0 GHz OH 
masers, producing a velocity separation of 0.018 km s$^{-1}$ mG$^{-1}$. 
The maximum magnetic field measured by \citet{C04} was 10.7 mG, with no 
single polarity spots being found, but given their delicate nature these 
masers might also not support large Zeeman shifts. It might be worth 
searching again for these and other excited OH masers to see what size 
fields they occur in.

\section{Conclusion} \label{sec:con}
The first examples of Zeeman splitting in 4.7 GHz exOH masers are 
presented here. We have determined reliable values for the 
velocity shifts due to the Zeeman effect, but because the value of 
the Land\'{e} g-factor is poorly known, we cannot determine an 
accurate value for the magnetic field. Perhaps these observations 
will spur someone to look at this problem in more detail and provide 
a better value. Because we have used the historical estimate of the 
conversion factor, the strength of magnetic fields associated with 
these Zeeman shifts are larger than have been measured previously 
in star-forming regions. As an example, if the conversion factor is 
a factor of seven smaller than assumed here, measured fields would 
have strengths of $B = 19.6$ mG for G173, and between 11.7 and 19.4 
mG for Mon R2. These are closer to currently estimated values of the 
$B$ field in star-forming regions, and would be even closer if the 
conversion factor was a factor of 10 smaller than the value used in 
Table \ref{tab:delta}. 

The linear polarization seen previously in Mon R2 is probably also 
due to the presence of a magnetic field, rather than being due to 
amplification of a background source, as was suggested in 
\citet{SCH98}. Time-varying linear polarization has been found in 
Mon R2 which will be reported separately.

The Stokes $V$ component displays a sine-shaped profile over the whole 
width of the maser line, as does the straight-line fit of the $V/I$ 
curve. These two features identify the circular polarization splitting 
as being due to the Zeeman effect. This is not nearly so obvious when 
working with the RCP and LCP components.

 It is important to have properly calibrated receivers for these 
measurements, otherwise, as pointed out by \citet{Wei62}, a component 
of $I$ will be present in the $V$ spectrum. This $I$ component can be 
removed but it is imperative that the fitted $I$ component does not 
have a velocity offset that will distort the $V$ signal. Because we 
are using a fully calibrated Mueller matrix the phase and amplitude 
calibration of our circular modes are established automatically.

Currently estimated strengths of magnetic fields in star-forming 
regions are based mainly on observations of Zeeman splitting in masers 
from the $^2\Pi_{3/2}$ levels of OH. These masers are limited in the 
range of field strengths they can measure because the circular components 
become too far apart for velocity coherence to occur in both modes.
The 4.7 GHz OH masers provide a method of measuring stronger fields, 
but at this stage the poorly known value of the Land\'{e} g-factor 
limits the accuracy that can be achieved.

\begin{acknowledgments}
The referee is thanked for their critical reading of the manuscript and 
helpful comments that have improved the text. This material is based upon 
work supported by the National Radio Astronomy Observatory and Green Bank 
Observatory which are major facilities funded by the U.S. National 
Science Foundation operated by Associated Universities, Inc. The 
observations were part of GBT projects AGBT22B-354, AGBT24A-429 and 
AGBT24B-513. 
\end{acknowledgments}

%

\vspace{5mm}
\facilities{GBT:100m}

\software{GBTIDL \citep{MGB13}, Matplotlib \citep{Hunter:2007}}

\appendix
\section{Slope of $V/I$ for Gaussian profiles}\label{appendix:slope}
Consider a single Gaussian  
\be
G(v) = A \me^{-(v - \mu)^2/2\sigma^2}  
\ee
where $A$ is the amplitude, $\mu$ is the central velocity of the profile 
and $\sigma$ is the standard deviation. To simplify the arithmetic in our 
derivations, the velocity scale is shifted so that $\mu = 0$.

Now Zeeman shift the RCP and LCP components by an amount $\delta$ in 
opposite directions, keeping the amplitudes of the two components the 
same. The $V$ = RCP -- LCP and $I$ = RCP + LCP components can be 
written as
\be
V = A\left[ \me^{-(v - \delta)^2/2\sigma^2} - \me^{-(v + \delta)^2/2\sigma^2} 
\right] \quad \mbox{ and } \quad I = A\left[ \me^{-(v - \delta)^2/2\sigma^2} 
+ \me^{-(v + \delta)^2/2\sigma^2} \right] \ . 
\ee

In principle, the amplitudes of RCP and LCP profiles need not be equal. 
For example, if the velocity coherence along the maser column is different 
for the two modes, then one component could have a larger amplitude than 
the other. To cater for the situation with different amplitudes, rather 
than introduce two amplitudes, $A$ is used as the amplitude of the RCP 
component, while the LCP amplitude is parameterised using a factor $f$ 
such that the LCP amplitude is $Af$. The $V$ and $I$ Stokes components 
can then be written as
\be 
V = A\left[ \me^{-(v - \delta)^2/2\sigma^2} - f\me^{-(v + \delta)^2/2\sigma^2} 
\right]  \quad \mbox{ and } \quad I = A\left[ \me^{-(v - \delta)^2/2\sigma^2} 
+ f\me^{-(v + \delta)^2/2\sigma^2} \right] \ .  
\ee

The arguments of the exponential terms can be expanded, giving
\be 
\frac{(v - \delta)^2}{2\sigma^2} = \frac{v^2}{2\sigma^2} - \frac{v\delta}{\sigma^2} + 
\frac{\delta^2}{2\sigma^2} \quad \mbox{ and } \quad \frac{(v + \delta)^2}{2\sigma^2} = 
\frac{v^2}{2\sigma^2} + \frac{v\delta}{\sigma^2} + \frac {\delta^2}{2\sigma^2}  \ ,
\ee
so that the exponential terms can be written in the form
\be \me^{-(v-\delta)^2} = \me^{-v^2/2\sigma^2}\,\me^{v\delta/\sigma^2}\,
\me^{-\delta^2/2\sigma^2} \quad \mbox{ and } \quad \me^{-(v+\delta)^2} = 
\me^{-v^2/2\sigma^2}\,\me^{-v\delta/\sigma^2}\,\me^{-\delta^2/2\sigma^2} \ . 
\ee

Noting that each term in the expressions for $V$ and $I$ contain the terms 
$\me^{-v^2/2\sigma^2}\,\me^{-\delta^2/2\sigma^2}$, these terms can be taken 
outside the brackets and canceled, which leads to 
\be 
\frac{V}{I} = \frac{\me^{v\delta/\sigma^2} - f\me^{-v\delta/\sigma^2}}
{\me^{v\delta/\sigma^2} + f\me^{-v\delta/\sigma^2}} = \frac{1 - 
f\me^{-2v\delta/\sigma^2}}{1 + f\me^{2v\delta/\sigma^2}} \ . 
\ee

If $\delta/\sigma \ll 1$, and $v \leq \sigma$ then $v\delta/\sigma^2 \ll 1$ 
so that $\me^{-2v\delta/\sigma^2}$ can be expanded to first order in a Taylor 
series to get
\be 
\frac {V}{I} = \frac{1 - f\left(1 - \frac{2v\delta}{\sigma^2} \right)}
{1 + f\left(1 + \frac{2v\delta} {\sigma^2} \right)} = \frac{(1-f) + 
\frac{2fv\delta}{\sigma^2}}{(1+f) + \frac{2fv\delta}{\sigma^2}} = \frac{(1 - f) 
+ \frac{2fv\delta}{\sigma^2}}{(1 + f) \left[ 1 + \frac{2fv\delta}{(1+f)\sigma^2} 
\right]} \ . 
\ee 

Provided $2fv\delta \ll (1+f)\sigma^2$, $V/I$ can be approximated by\\
\be \label{eq:V/I} 
\frac{V}{I} = \left(\frac{1-f}{1+f} + \frac{2fv\delta}{(1+f)\sigma^2} \right) 
\left(1 - \frac{2fv\delta}{(1+f)\sigma^2} \right)  = \frac{1-f}{1+f} + 
\frac{4fv\delta}{(1+f)^2\sigma^2} \ . 
\ee

This is a linear function with a slope $m = 4f\delta/(1+f)^2\sigma^2$ that 
cuts the velocity axis at $v = (f-1)(f+1) \sigma^2/4f\delta$. It is 
straightforward to show that for $f=1$ 
\be
m = \delta/\sigma^2
\ee
and the $V/I$ line cuts the velocity axis at $v = 0$ . If there is a 
significant difference in the amplitudes, then both the slope and the 
$v$-axis intersection are affected.

The largest measured value of $f$ from Table~\ref{tab:RCP_LCP} is 1.009 
for G173 which is the observation with the smallest SNR. Substituting 
this value into the above formulae changes the velocity intersection 
from 0 to 0.0004 and the slope is affected by a factor $< 10^{-4}$. 
In both instances, these factors are less than the uncertainties 
associated with the derived values and therefore the difference in 
amplitudes is inconsequential.

\section{Derivative of Stokes $I$ proportional to Stokes $V$} 
\label{appendix:derivative}

From the definition of a derivative, approximate expressions for 
$dI/dv$ are given by
\be
\frac{dI}{dv} \approx \frac{I(v+\delta) - I(v)}{\delta} \quad\quad 
\mbox{ and } \quad\quad \frac{dI}{dv} \approx \frac{I(v) - I(v - 
\delta)}{\delta} \ ,
\ee
from which it follows that 
\be
 I(v+\delta) \approx I(v) + \delta\frac{dI}{dv}  \quad 
\mbox{ and } \quad I(v-\delta) \approx I(v) - \delta\frac{dI}{dv}.
\ee
If RCP and LCP profiles have equal amplitudes then
\be
\textrm{RCP} =  \frac{I(v-\delta)}{2} \approx \frac{I(v)- \delta 
\frac{dI}{dv}}{2}\quad \mbox{ and } \quad \textrm{LCP} = \frac{I(v 
+\delta)}{2} \approx \frac{I(v) - \delta \frac{dI}{dv}}{2} 
\ee
which gives
\be \label{eq:didv}
I(v) = \textrm{RCP + LCP}  \quad\quad \mbox{ and } \quad\quad V(v) 
= \textrm{RCP -- LCP} = -\delta \frac{dI}{dv} \ .
\ee
The above derivation requires equal amplitude profiles for RCP and LCP
and that $\delta$ is small, but sets no criteria for how small $\delta$
must be.

Using the two Gaussian maser profile model described in Appendix A, the 
derivative of the $I$ component with respect to velocity is 
\be \label{eq:dI/dv} 
\begin{split}
\frac{dI}{dv} &= \frac{A}{\sigma^2} \left[\delta\left( e^{-(v-\delta)^2/2\sigma^2} - 
fe^{-(v+\delta)^2/2\sigma^2} \right) - v\left( e^{-(v-\delta)^2/2\sigma^2} +
fe^{-(v+\delta)^2/2\sigma^2} \right) \right] \\ &= \frac{\delta}{\sigma^2}V 
- \frac{v}{\sigma^2}I \   
\\ &= V \left[ \frac{\delta}{\sigma^2} - \frac{v (1+f)^2}{(1-f^2)\sigma^2 + 
4fv\delta} \right] \ .
\end{split}
\ee
where the last equation has been obtained by replacing $I/V$ with 
Eq.~(\ref{eq:V/I}). Using this equation, the derivative of $I$ can be 
computed and fitted to $V$ to determine  solutions for both $\delta$ 
and $f$.

For $f = 1$ the derivative of the $I$ component simplifies to  
\begin{eqnarray} \label{eq:dI/dv_approx}
\frac{dI}{dv} & = & V \left[\frac{\delta}{\sigma^2} - \frac{1}{\delta} 
\right] \nonumber \\ 
& =  & -\frac{V}{\delta} \quad \mbox{for } \delta \ll \sigma \, 
\end{eqnarray}
which agrees with Eq.~(\ref{eq:didv}) with the additional criterion 
that $\delta \ll \sigma$.

\bibliographystyle{aasjournal}
\bibliography{4.7CircPol} 


\end{document}